# Step-Wise Computational Synthesis of Fullerene $C_{60}$ derivatives. 1.Fluorinated Fullerenes $C_{60}F_{2k}$

Elena F.Sheka

*Research Department, Peoples` Friendship University of the Russian Federation, Moscow*
sheka@icp.ac.ru

**Abstract.** The reactions of fullerene $C_{60}$ with atomic fluorine have been studied by unrestricted broken spin-symmetry Hartree-Fock (UBS HF) approach implemented in semiempirical codes based on AM1 technique. The calculations were focused on a sequential addition of fluorine atom to the fullerene cage following indication of the cage atom highest chemical susceptibility that is calculated at each step. The effectively-non-paired-electron concept of the fullerene atoms chemical susceptibility lays the foundation of the suggested computational synthesis. The obtained results are analyzed from energetic, symmetry, and the composition abundance viewpoints. A good fitting of the data to experimental findings proves a creative role of the suggested synthesis methodology.

## 1. Introduction

A fluorinated fullerene $C_{60}$ decade, started with the first synthesis in 1991 (see reviews [1-4]) and ended by a generalizing theoretical approach to the fluorinated fullerenes characterization in 2003 [5], without any doubt is one of the best examples of a power and ability of the modern chemistry to produce, characterize, and describe at microscopic level a new family of fascinating chemicals. It is especially impressive since the number of species, hidden under the general formula $C_{60}F_{2k}$ (with $k=1,\ldots, 30$), is definitely countless if all possible isomers at each particular $k$ are taken into account. The first breaking through the problem has been made by experimentalists who showed that 1) species with even number of fluorine atoms could be observed only, 2) not all $k =1,\ldots, 30$ fluorinated products but only a restricted set of them could be produced and identified in practice, and 3) a very limited number of isomers, from one to three, were revealed. Thus, products with chemical formula $C_{60}F_{18}$, $C_{60}F_{36}$, $C_{60}F_{48}$ dominate in the production list whilst minor products from $C_{60}F_2$ to $C_{60}F_{20}$ have also been identified [6]. Mass spectrometry, IR, $^{19}F$ and $^3He$ NMR spectroscopy manifested themselves as reliable constituents of a convincing analytical platform for the species identification.

At the same time, quantum chemical (QCh) simulations have faced the many-fold isomerism problem in the full measure. Thus, to suggest a convincing atomistic structure of the products produced experimentally, one has to make choice of one to three isomers [6] among 600 873 146 368 170 isomers of $C_{60}F_{36}$ [7] and 23 322 797 475 ones of $C_{60}F_{48}$ [7]. To make computations feasible one has to restrict the isomer number to units. Obviously it might be possible if some global regularities that govern fluorination process can be exhibited. To start going on the way, it was necessary to answer the following questions:

1. To what kind of chemical reactions does the fullerene fluorination belong?
2. What is the pathway for fluorination of fullerene $C_{60}$ and does fuorination of $C_{60}F_2$ to, say, $C_{60}F_{48}$ follow a single pathway and to take place in regular steps?
3. How target carbon atoms of the fullerene cage have to be chosen priori to any subsequent step of fluorination?
4. How can these measures restrict the number of possible isomers?

Answering the first question, the computing community has considered the fullerene fluorination as a radical reaction [8] of one-by-one addition of fluorine atoms where the first addition drastically violates a double C-C bond of the fullerene cage while the next addition completes the transformation of the bond into a single one.

Experimental data, particularly the exhausted study of electronic structure and chemical bonding of fluorinated fullerenes by XPS, NEXAFS, UPS, and vacuum-UV absorption [9] give a positive answer to the second question so that fuorination of $C_{60}F_2$ to $C_{60}F_{48}$ follows a single pathway and occurs in regular steps.

As for the choice of target atoms for the subsequent steps, a computing scientist had to solve dilemma of either to accept a full equality of the fullerene $C_{60}$ atoms with respect to chemical reactivity or to look for regioselectivity of the cage atoms. Obviously, the first suggestion, which has been accepted by the majority of the computational community until now, was absolutely impotent in the solution of the above isomerism problem. So that to proceed with the problem solution, one has to accept a particular regioselectivity of fullerene atoms. The first suggestion concerns the selection between 6,6 double bonds joining two hexagons and 6,5 single bonds framing pentagons in favor of the former. The next suggestion deals with the separation between carbon sites of the subsequent addition of fluorine atoms either via bonds (1,2-addition, adjacent carbon sites) or via space (1,3- and 1,4-additions, widely spaced carbon sites). Matsuzawa et.al. [10] were first to consider the point proving that 1,2-addition is more preferable for fluorine. In a decade, the conclusion was confirmed by Jaffe [5]. The next question concerns the succession of the 1-2-additions. Following a concept of a supposed increasing of aromaticity of the fullerene hexagons caused by fluorine addition generalized by Taylor [4], a contiguous $F_2$ addition was suggested where the preference is done to C-C double bonds adjacent to the prior addition sites.

The above issues are quite important for designing model structures but they should be nevertheless qualified as general recommendations only. The main problem concerning the number of isomers attributed to each of the $C_{60}F_{2k}$ species still remained. The way out of the situation was suggested by Clare and Kepert [11] who proposed to restrict the isomer number by those possessing a three-fold symmetry axis. The suggestion was based on a clear experimental evidence of the $C_{3v}$ symmetry of the crown-shaped $C_{60}H_{18}$ species [12]. That has been spread over $C_{60}F_{18}$ as well [11] and later on has been expanded over $C_{60}F_{36}$ and $C_{60}F_{48}$ adducts also [13-16]. However, under these assumptions, the number of possible isomers still retains quite large and constitutes, say, 2695 for $C_{60}F_{36}$ species [17]. The next step towards decreasing the isomer number has implied concrete suggestions concerning the molecules symmetry. Thus, a crown-shaped structure of $C_{60}F_{18}$ points to the $C_{3v}$ symmetry [11], the consideration of $T$, $C_3$, $S_6$, and $D_{3d}$ structures for $C_{60}F_{36}$ has allowed for decreasing the isomer number to 63 [14], 94 isomers have been selected in the case of $C_{60}F_{48}$ [13]. The justification of the assumption has been mainly based on the $^{19}F$ NMR spectra of $C_{60}F_{18}$ [18], $C_{60}F_{36}$ [19], and $C_{60}F_{48}$ [20, 21]. However, a thorough analysis of both $^{19}F$ NMR and 2d COSY spectra has shown that the structure assignment, based on the theoretical predictions, is not absolutely secure, particularly for $C_{60}F_{36}$ and $C_{60}F_{48}$ species (a detailed discussion see in Section 5.4). Therefore, the problem of the assignment of the most stable isomers can not be considered as definitely solved.

An alternative approach to the problem can be suggested on the basis of the inherent regioselectivity of 60 pristine fullerene atoms. The latter is based on a partial radicalization of the $C_{60}$ molecule caused by the appearance of *effectively unpaired electrons* (EUPEs) due to weakening interaction of its odd electrons [22-27]. These electrons, distributed over the cage atoms, enhance the atom chemical reactivity providing a numerically definite *atomic chemical susceptibility* (ACS). Due to non-monotonic EUPEs distribution, the ACS mapping over cage atoms readily highlights target atoms that are characterized by the highest ACS. Thus taking ACS as a quantitative pointer of the readiness of each atom to enter the chemical reaction, one is able to make a definite choice of targets at each stage of the reaction and, consequently, to perform a step-wise computational synthesis of fullerene derivatives of any composition. Exemplifying by the first steps of the fullerene fluorination [25-27], the approach is turned out to be a proper tool for a step-wise synthesis of the halogenated $C_{60}(Hal)_{2k}$, hydrogenated $C_{60}H_{2k}$, aminated $C_{60}(Am)_m$ fullerene adducts in a predictable manner. The current paper presents results related to the $C_{60}F_{2k}$ family. Hydrogenated and aminated families are discussed elsewhere [28, 29].

## 2. Methodology of computations

A traditional theory of chemical bonding has taken conceptual and quantitative determination in terms of either bond [30] or Wiberg [31, 32] indices within the framework of the single-determinant close shell restricted Hartree-Fock (RHF) approximation. Addressing to odd-electron systems (the term indicates that the number of valence electrons of each fullerene atom is larger by one than that of interatomic bonds formed by the atom), this corresponds to a limit case of strong coupling between the electrons. This requirement is met in the case of ethylene and benzene where odd electrons are fully covalently coupled in form of π electrons. However, even for naphthalene, not saying about higher acenes [33, 34], fullerenes [22-27], carbon nanotubes CNTs) [33, 35], and graphene [36] it is not the case due to enlarging length of C-C bonds compared to ones of the benzene molecule which causes a noticeable weakening of the interaction between the electrons so that a part of odd electrons are excluded from the covalent coupling and become effectively unpaired.

The approach generalization for systems with weakly interacting electrons ultimately requires taking into account the electrons correlation and passing to computational schemes that involve full configurational interaction (CI). However, the traditional complete-active-space-self-consistent-field (CASSCF) methods that deal correctly with two electron systems of diradicals and some dinuclear magnetic complexes, cannot handle systems with a large number of the electrons due to a huge number of configurations generated in the active space of the system so that for $m$ singly occupied orbitals on each of $n$ identical centers $2^{mn}$ Slater determinants should be formed by assigning spins up or down to each of the $nm$ orbitals [37]. It has been accepted until recently, that CASSCF type approaches are non-feasible for many-odd electron systems such as fullerenes, CNTs, and graphene. Thus, addressing single-determinant approaches appeared to be the only alternative.

The open-shell unrestricted broken spin-symmetry (UBS) approach suggested by Noodleman [38] is well elaborated for both wave-function and electron-density QCh methodologies, based on unrestricted single-determinant Hartree-Fock scheme [39] (UBS HF) and the Kohn-Sham single Slater determinant procedure DFT (UBS DFT) [40]. The UBS approach main problem concerns spin-contamination of the calculation results. The interpretation of UBS results in view of their relevance to physical and chemical reality consists in mapping between the eigenvalues and eigenfunctions of exact and model spin Hamiltonians. The implementation of UBS HF approach, both *ab initio* and semiempirical, is standard and the wished mapping is quite straightforward. Recently appeared the first attempts of many-body configurational interaction (CI) calculations of polyacenes [34] and graphene [41] have provided a strong support in favor of the UBS HF approach [36] and highlighted a high ability of the latter to quantitatively describe practically important consequences of weak interaction between odd electrons of nanocarbons. A semiempirical UBS HF calculation scheme based on AM1 approach was used in the current study and providing its high efficacy only, the performance of massive calculations concerned with the isomerism of the $C_{60}F_{2k}$ family has been succeeded.

3. Synopsis of the values under determination

EUPEs provide a partial radicalization of the species and, thus, a considerable enhancement of its chemical reactivity. Since EUPEs are produced by spin-contaminated UBS solutions they are directly connected with the spin contamination

$$C = \langle \hat{S}^2 \rangle - S(S+1), \qquad (1)$$

where $\langle \hat{S}^2 \rangle$ is the expectation value of the total spin angular momentum that follows from the UBS solution. Actually, as shown in [42], the total EUPEs number $N_D$ is expressed as

$$N_D = 2\left( \langle \hat{S}^2 \rangle - \frac{(N^\alpha - N^\beta)^2}{2} \right) \qquad (2)$$

where $N^\alpha$ and $N^\beta$ are the number of electrons with spin α and β, respectively, and $N^\alpha - N^\beta$ determines the spin multiplicity. On the other side, the spin contamination produces an extra spin density (that is particularly evident for singlet state) so that $N_D$ is expressed as a trace of the density [43]

$$N_D = trD(r|r'). \qquad (3)$$

Therefore, to quantify $N_D$ one has to know either $\langle \hat{S}^2 \rangle$ or $trD(r|r')$.

For a single-Slater-determinant UBS HF function, the evaluation of both quantities is straightforward since the corresponding coordinate wave functions are subordinated to the definite permutation symmetry so that each value of spin $S$ corresponds to a definite expectation value of energy [44].

Thus, $\langle \hat{S}^2 \rangle$ is expressed as [45]

$$\langle S^2 \rangle = \frac{(N^\alpha - N^\beta)^2}{4} + \frac{N^\alpha + N^\beta}{2} - \sum_{i,j=1}^{NORBS} P_{ij}^\alpha * P_{ij}^\beta. \qquad (4)$$

Here $P_{ij}^{\alpha,\beta}$ are matrix elements of electron density for α and β spins, respectively. Similarly, Ex.(3) has the form [25]

$$N_D = \sum_{i,j=1}^{NORBS} D_{ij}, \qquad (5)$$

where $D_{ij}$ are matrix elements of the spin density expressed as [25]

$$D = \left(P^\alpha - P^\beta\right)^2. \qquad (6)$$

The latter expression is related to the NDDO approximation that lays the foundation of AM1/PM3 semiempirical computational schemes. The summation in (5) is performed over all atomic orbitals.

The atomic origin of the UBS HF function produces another important relation concerning the partitioning of the $N_D$ value over the system atoms

$$N_D = \sum_A^{NAT} N_{DA}, \qquad (7)$$

where $N_{DA}$ is expressed as [25]

$$N_{DA} = \sum_{i \in A} \sum_{B=1}^{NAT} \sum_{j \in B} D_{ij}, \qquad (8)$$

and is attributed to the EUPEs number on atom *A* (the summation in (7) and (8) is performed over all atoms). Since EUPEs themselves are a "quality index" of the enhanced reactivity, $N_D$ and $N_{DA}$ quantities just quantify the index representing *molecular* and *atomic chemical susceptibility* (ACS), respectively. A correct determination of both values is well provided by AM1/PM3 UBS HF solution [25] of the CLUSTER-Z1 software [46] used in the current study.

## 4. Algorithm of computational synthesis of fullerenes derivatives

The analysis of chemical activity of both carbonaceous and siliceous fullerenes presents the first practical implementation of the concepts described above [22-24]. From the EUPEs viewpoint, the $C_{60}$ and $C_{70}$ molecules are characterized by a partial exclusion of the odd electrons from the covalent bonding that results in EUPEs constituting ~10%-odd-electron fraction, while all odd electrons of the $Si_{60}$ molecule are not paired providing the 60-fold radicalization of the molecule. The finding highlights evident reasons for the failure in producing the species on practice. ACS ($N_{DA}$) maps complete the structural description of the $C_{60}$ and $C_{70}$ molecules by their "chemical portraits" [25-27]. According to the latter, the $C_{60}$ molecule consists of six identical naphthalene-core fragments forming a $6*C_{10}$ configuration.

The chemical portrait of the $C_{60}$ molecule is shown in Fig.1a and b while ACS map of the molecule atoms belonging to groups 1-5 is presented in Fig.1c. According to the figure, the initial step of any addition chemical reaction involves atoms of the highest ACS of group 1. There are 12 identical atoms that form 6 short C-C bonds belonging to 6 identical naphthalene-core fragments. We may choose any of the pairs to start the reaction of attaching any addend to the fullerene cage, including fluorine atoms and/or molecules. When the first adduct $C_{60}R_1$ is formed, the reaction proceeds around the cage atoms with the biggest $N_{DA}$ values resulting in the formation of adduct $C_{60}R_2$. A new ACS map reveals the sites for the next addition step and so on. The reaction stops when all $N_{DA}$ values are fully exhausted. Following these methodology, a complete list of fluorinated fullerenes $C_{60}X_{2k}$ has been synthesized.

## 5. Results and discussion

### 5.1. Start of the $C_{60}$ fluorination.

When starting fluorination of $C_{60}$, one fluorine molecule is placed in the vicinity of the selected atoms of group 1 (31 and 32 in the case) (Fig. 2a) and a full optimization of the complex geometry in the singlet state is performed. As occurred, the fluorine molecule is willingly attached to the cage, however, two adducts are possible depending on the fluorine molecule orientation with respect to the chosen C-C bond. If the molecule axis is parallel to the bond, adduct $C_{60}F_2$ (**I**, Fig. 2b) is formed. If the molecule axis is even slightly inclined towards the bond a complex $C_{60}F_1 + F_1$ (**II**+ $F_1$, Fig. 2c) is obtained.

Fig.3. presents the ACS maps of pristine $C_{60}$ cage and the cage after formation of adducts **I** and **II** following the atom numeration in the output file. Crosses mark initial target atoms 31 and 32. As seen from the figure, attaching either one or two fluorine atoms changes the initial map considerably and differently in both cases. When two atoms are attached to the cage, the $N_{DA}$ values become zero for target atoms 31 and 32 and star-marked atoms 18, 20, 38 и 55 become the most active (Fig. 3a). When one atom is attached, remaining target atom 31, which is adjacent to the first target atom 32, dominates on the adduct ACS map (Fig. 3b). The picture clearly evidences a readiness of the $C_{60}$ cage to complete the reaction by adding fluorine atom to atom 31. Following this indication, and keeping configuration of the **II**+F complex, we add the second fluorine molecule, as shown in Fig. 2d. In due course of the structure optimization, a new adduct $C_{60}F_2 + 2F$ (**III**+2F, Fig. 2e) is formed. Geometry and electronic properties of adduct **III** are fully identical to those of adduct **I** that is confirmed by a complete identity of their $N_{DA}$ maps as well. Therefore, independently of either one-stage (Fig. 2b), or two-stage (Fig. 2e) processes of the $F_2$ attachment to the fullerene cage occur, the same final adduct $C_{60}F_2$ is formed. Obviously, two-stage reaction should prevail in practice. It is radical by nature and follows a qualitative scheme suggested by Rogers and Fowler [47].

The next step of the reaction is governed by the predominance of atoms 18, 20, 38 and 55 on the ACS map of $C_{60}F_2$ (see Fig. 3a) over other atoms. These atoms form two identical pairs of short C-C bonds located in the equatorial plane with respect to first two target atoms (see lightly colored atoms in Fig. 4a). One of these pairs is taken as targeting and the procedure of attaching $F_2$ to the pair atoms repeats that described above. Consequently, a molecule $C_{60}F_4$ is formed (Fig. 4b). The ACS map is calculated for the product to select target atoms for the next attaching. As seen from Fig.4 a subsequent $F_2$ addition is not contiguous (see Fig.4c) as was suggested by Taylor [6].

From the computational viewpoint, the result of one-atom addition that occurred at each stage of the two-stage process of the $F_2$ addition does not depend on which namely, molecular or atomic addend attacks the cage. That is why a subsequent addition of fluorine atoms one-by one will be considered in what follows as a series of subsequent steps, consisting of two stages which involved calculations of two adducts related to two reactions $C_{60}F_{2к}+F= C_{60}F_{2к+1}$ and $C_{60}F_{2к+1}+F = C_{60}F_{2(к+1)}$, k=1, 2...30. Each step is controlled by the fullerene cage ACS map of the preceding adducts, namely, $C_{60}X_{2к}$ and $C_{60}F_{2к+1}$, respectively. Actually, when the difference between the high-rank $N_{DA}$ values is not quite pronounced, every step is additionally complicated by expanding calculations over a restricted set of isomers which are pointed out by a set of high-rank values on the ACS map. A final choice of the most stable species is subordinated therewith to the preference of the structure with the least total energy.

### 6.2. $C_{60}F_2$-$C_{60}F_8$ adducts

Let us consider the synthesis of fluorinated adducts in the working regime exemplifying the procedure for $C_{60}F_2$-$C_{60}F_8$ adducts. Chart 1 presents high-rank $N_{DA}$ values of the species from $C_{60}$ to $C_{60}F_8$. The $N_{DA}$ data are ordered from the biggest to the lowest one and only a small part of high-rank data is shown. Fluorinated adducts are marked FN.

Cage atom 32 was chosen as the first one to be attacked by fluorine. Atom 31 heads the list of the $N_{DA}$ data of adduct F1 and obviously points to the place of the next attack. The head of the $N_{DA}$ list of F2 involves to pairs one-bond-connected atoms 38&55 and 20&18 (they are shown in Fig.4a. The pairs are fully equivalent and as shown by calculations, adducts series started from each of them are equivalent as well. We proceed with atom 38 to form F3, whose $N_{DA}$ list immediately highlights the pairing atom 55. The $N_{DA}$ list of F4 is open by one-bond-connected atoms 42&48 and includes atom 22 pair of which is shifted in the depth of the list. Continuing addition by attacking atom 42, we obtain F5 with atom 48 possessing the highest $N_{DA}$. The $N_{DA}$ list of F6 is typical for the majority of succeeding addition events and is headed by two (sometimes three atoms) with comparable $N_{DA}$ values while the corresponding one-bond-connected atoms are shifted in the depth of list. Two such pairs of $F_6$ are shown by different coloring in Chart 1. This is a typical case when a study of a few isomers is needed. The first isomer F7-1 corresponds to attacking atom 60 and highlights atom 59 for the next attack. The second isomer F7-2 starts on atom 58 and exhibits its pairing atom 57. Two F8 adducts followed from these two F7 isomers differ by the total energy that is 517.162 and 517.216 kcal/mol for isomers F8-1 and F8-2, respectively. In spite of seemingly small difference in the energy, favoring isomer F8-1 occurred quite reliable. Checking series from F8-1 and F8-2 isomers showed that all subsequent species are much higher total energy in the second case. Atomic view of the F2K species for $k$=1, 2, 3, and 4 are shown among the others in Fig. 5. As seen from the figure, neither F4 nor F6 and F8 follow the contiguous scheme of the addition suggested by Taylor [4]. Not hypothetical growing of the aromaticity of adjacent bonds but the redistribution of the EUPEs density over the cage atoms is the governing factor of the reaction pathway.

Total energy of adducts as well as their main geometric parameters are presented in Table 1. The latter concern C*-F and C*-C* bond lengths where C* marks cage atoms bound to a fluorine one. Changing the total energy in due course of fluorination is shown in Fig.6 as function of the coupling energy $E_{cpl}$ needed for the addition of every next pair of fluorine atoms from the pair number $k$. Supposing that the reaction occurs in the gaseous fluorine, the energy is determined as

$$E_{cpl} = \Delta H_{2k} - \Delta H_{2(k-1)} - \Delta H_{F_2}. \qquad (9)$$

Here $\Delta H_{2k}$ and $\Delta H_{2(k-1)}$ are heats of formation of F2K and F2(K-1) products while $\Delta H_{F_2}$ is the heat of formation of fluorine molecule equal to -22.485 kcal/mol. The data lay within {-91; -83} kcal/mol interval demonstrating a tight binding of fluorine with the fullerene cage. The corresponding best values for F2 and F4 obtained by Matsuzawa et al [10] in the close-shell RHF AM1 approximation and corrected by $\Delta H_{F_2}$ are practically constant and constitute –63.5±0.2 kcal/mol. The data increase in about 1.5 times when going from RHF to UHF approximation. Close-shell B3LYP/4_31G calculations at B3LYP/STO-3G geometry result in the correspondingly $\Delta H_{F_2}$-corrected values of -92.4 and -101.7 kcal/mol for F2 and F4 species, respectively [5] that are close to the UHF results.

As seen from Table 1, the total energy remarkably decreases when fluorination proceeds, that favors polyaddition. Actually, this occurs experimentally and $C_{60}F_{18}$, $C_{60}F_{36}$, and $C_{60}F_{48}$ are usually the main products obtained [4]. As shown by groups of Taylor [4] and Boltalina [2], the fluorination yield greatly depends on the reaction conditions. In fluorination by fluorine gas, the EI mass spectrum showed a continuous spectrum of derivatives from $C_{60}F_2$ to $C_{60}F_{42}$. However, it was unclear if the lower fluorinated species were compounds or merely fragmentation ions. And only after particular measures were taken, the authors succeeded in separating lower fluorinated species $C_{60}F_2$, $C_{60}F_4$, $C_{60}F_6$, and $C_{60}F_8$ [48, 49] confirming adducts stability.

The top part of Fig.6 presents changing of the EUPES total number as a function of the number of fluorinated pairs added to the fullerene cage. As a total, the value gradually decreases when fluorination proceeds, thus, logically molecular chemical susceptibility $N_D$ behaves as a "pool of chemical reactivity" that is being worked out in due course of the reaction proceeding. At the very beginning at low $k$, $N_D$ slightly increases due to increasing the number of elongated C-C bonds caused by the cage structure $sp^2$-$sp^3$ reconstruction under fluorination.

### 5.3. $C_{60}F_{10}$-$C_{60}F_{18}$ adducts

A detailed procedure of computational fluorination of $C_{60}$ cage discussed above is conserved through over a full cycle of $C_{60}F_2$ to $C_{60}F_{60}$ fluorination. A concise synopsis related to the formation of products from $C_{60}F_{10}$ to $C_{60}F_{18}$ that required a study of 11 isomers is presented in Chart 2. The content of each cell involves the isomer name, the cage atom numbers, which were taken from the high rank part of the $N_{DA}$ list of the preceding isomer and addition to which provided the current isomer formation, and the isomer total energy. Isomers with the least total energy are bold marked. When reading out the chart, the following picture is developped. Starting with isomer F8-1 and looking at the $N_{DA}$ list in chart 1, one can obtain two isomers F10 by adding a pair of fluorine atoms to either atoms 22 and 33 (F10-1) or atoms 23 and 24 (F10-2). Total energies of the two isomers are of 414.700 and 415.812 kcal/mol. The energy difference favors the first one which we put at the head of $C_{60}F_{10}$ - $C_{60}F_{18}$ series in chart 2. The high-rank part of the list of F10-1 involves three pairs which originate three isomers F12 shown in chart 2. Isomer F12-2 possesses the least energy and fluorine atom addition to its high-rank atoms produces two isomers F14. Choosing isomer F14-1, we proceed with two isomers F16, among which isomer F16-2 is more energetically favorable. Adding fluorine atoms to atoms 23 and 24 of isomer F16-2, we obtain isomer F18-1. Shown for comparison in chart 2 is presented the case when adduct F18 is produced from isomer F16-1. As seen from the chart, thus formed F18-2 and F18-3 adducts posses much bigger energy and are energetically non profitable with respect to adduct F18-1.

Equilibrated structures of adducts from the $C_{60}F_{10}$ - $C_{60}F_{18}$ series which correspond to those of the least total energy are shown in Fig. 5. Drawn in the same projection, the pictures allow vividly exhibiting a consequent deformation of the fullerene cage caused by fluorination until F18 is formed. The last row of pictures in Fig.5 presents F18 in different projections to highlight its crown-like structure of $C_{3v}$ symmetry well known experimentally [50-52]. Total energy and geometric parameters of adducts are given in Table 1. Data plotting in Fig.6 show that both $N_D$ quantity and coupling energy differ not so much within the series and retain close to the biggest values just disclosing high chemical reactivity of individual adduct and strong tendency to further fluorination.

### 5.4. $C_{60}F_{20}$-$C_{60}F_{36}$ adducts

Continuing the computational procedure, a series of $C_{60}F_{20}$-$C_{60}F_{36}$ adducts were produced. About 20 isomers were considered to make choice of the most stable isomers. Such selected isomers are shown in Fig.7 while their total energy and geometric parameters are given in Table 1. Oppositely to the previous series, the current one does not ended by adduct of high symmetry. The obtained F36 adducts has $C_1$ symmetry.

$C_{60}F_{36}$ is one of the most studied among other fluorinated fullerene species. However, since the first recording of the substance [53] till the last one [54] there still has not been clear vision of the composition and symmetry of the species. A big temptation to see a direct connection between F18 and F36 [6] influenced looking for $C_3$-based structures of F36 both computationally [14-17] and experimentally [52, 54-57]. However, all experimental findings occurred to be of complicated structure and cannot be interpreted from the high symmetry position. This led to conclusion about a complicated isomer composition of the produced material which involves isomers of $C_3$, $C_1$, and $T$ symmetry. Another view on the problem might be based on the computed F36. This concerns interpreting experimental findings not from a rigid "yes/no" symmetry viewpoint ($C_1$ or other symmetry), but implies the symmetry "grayness" or, by other words, its "expansion" over a set of high symmetry contributions, different with respect to different physical properties [58, 59]. One cannot exclude that in this case fitting experimental and computational data could be quite satisfactory as this is in the case of $C_{60}$, whose symmetry is not a strict $I_h$ but gray $C_i$ [60].

As seen from Fig.6, fluorination F20 to F36 covers the main step when the pool of the cage chemical reactivity has been being worked out. The remaining $N_D$ values become small. Simultaneously, a considerable decreasing of the coupling energy by absolute value takes place. Both factors evidently point to weakening the reaction ability that evidences the reaction termination in a short time.

### 5.5. $C_{60}F_{38}$-$C_{60}F_{48}$ adducts

Due to smallness of the $N_D$ value, the next steps of the addition reaction require higher attention and thorough investigation of a number of isomers at each step. Nevertheless, trying a few roots of the reaction continuation, the same set of energetically stable isomers has been obtained. The species are presented in Fig. 8. Their total energy and geometric parameters are given in Table 1. As seen from Fig.8, the structure of adducts becomes visually more and more symmetric when fluorination proceeds. However, the point symmetry of F48 is $C_1$.

Analyzing Fig.6 for this stage of reaction, one should conclude that experimental realization of high fluorinated $C_{60}$ is approaching the end since 1) the chemical reactivity pool is worked out and 2) coupling energy actively decreases approaching to zero. Therefore, F44-F50 species must be final ones in the row of fluorinated $C_{60}$ since coupling energy of F52 and higher species becomes positive. One of the first report on producing F48 [61] states that the substance can be obtained only in due course of long-time fluorination and that depending on the temperature one can obtain either F46 or F48 species. Both species are due to the polyaddition realization inherent to fluorination process. Just stopping the reaction afterwards allows for accumulating the species in a considerable amount. This makes the production of F48 more favorable than either F18 or F36 in spite of much worse conditions for the reaction occurrence concerning $N_D$ values and coupling energy. As for higher fluorinated fullerenes, no indication about their observation is known [6] besides F60 which supposedly was observed as trace amounts in $^{19}$F NMR study [6].

F48 is the main target of the stage on which a lot of efforts were concentrated. The chemical formula of F48 intuitively promises high symmetry if $C_{3v}$ symmetry of F18 is taken into account. The first $^{19}$F NMR analysis of the species showed [61] that when NMR spectrum of F46 was very complicated that one of F48 demonstrated clear patterns characteristic for high symmetry. It was suggested that two chiral isomers of $D_3$ symmetry formed the studied sample. All further investigations were aimed at proving high symmetry of the species, both computationally [13, 16] and experimentally [57, 62-64]. In spite of heavy efforts undertaken to clarify the symmetry problem, there are still features which do not fit in the high symmetry suggestion. Those are extra dots on 2D COSY $^{19}$F NMR diagramme [61], non sufficient accuracy of gas electron diffraction due to a great number of parameters used under interpreting the data obtained [63]. The latter can be addressed as well to high-resolution X-ray powder diffraction [64] so that only a relative conclusion could be made that, say, $D_3$ symmetry of the molecule fits experimental data under applied conditions better than $S_6$ or $T_d$. Under these conditions it was interesting how precisely may fit experimental data the structure shown in Fig.8. Dr.R.Papoular from the Leon Brillouin Laboratory from CEN-Saclay agreed to perform Rietveld refining of his high-resolution X-Ray powder diffraction structure data for $C_{60}F_{48}$ powder by using molecular structure F48 followed from the current study [65]. Fig.9 presents the refining results for the molecule in comparison with those obtained for $D_3$-symmetrical species. The latter was accepted as the best fitting when comparing molecular structure of $D_3$, $S_6$, and $T_d$ symmetry [64]. As seen from the figure, the fitting for $C_1$ molecule visually is not only not worse than that of $D_3$ symmetry but even better in the region of the first most intensive peak. Table 2 summarises numerical comparison of the two Rietveld refinements added by the data for $S_6$ molecule. It is possible to conclude that the $C_1$ molecule refinement is between $D_3$ and $S_6$ ones that is why if we did not know the molecule symmetry we should have to conclude that it is either $D_3$ or $S_6$. At any rate, the molecule had to be attributed to high symmetry, not to $C_1$. This analysis shows how ambiguous might be conclusions made on the basis of experimental data.

As for the discussed X-Ray diffraction case, the Rietveld refinement started from the modelling crystal packing. This procedure involves a large set of approximations (6-exp-1 intermolecular potential, a particular algorithm of the Coulomb energy calculation, etc.) that make the refining procedure to be dependent on numerous parameters. Not all of them are physically real. Thus, when calculating the Coulomb energy within the bond-center charge model it was assumed [64] the charges -0.086 and -0.380 $e$ on atoms C and F of each C-F bond, respectively, and 0.466 e on the charge center in the bond middle, while the unfluorinated carbons had no charge. This assumption contradicts drastically to the real picture of the charge distribution over the molecule atoms, which is shown in Fig.10. F48 species is presented in the figure by $D_3$ (curve with open circles) and $C_1$ (histogram) molecules. The structure of the former was taken as suggested in [65]. This very $D_3$ molecule was used for the Rietveld refinement in [64]. As seen from the figure, the general pattern of the charge distribution for both molecules is well similar. The difference between two plottings just reflects the difference in the molecule symmetry. In both cases, fluorinated carbon atoms form three main groups with charges of the first group within 0.135-123, $e$ ($C_1$) (a spike at 0.185 is a direct evidence of low symmetry) and 0.132-0.125, $e$ ($D_3$) interval; of the second group within 0.087-0.073, $e$ ($C_1$) and 0.089-0.080, $e$ ($D_3$); and of the third group -0.098- -0.128, $e$ ($C_1$) and -0.119, $e$ ($D_3$). The first group involves atoms one neighbour of which is not fluorinated. All atoms of the second group are fluorinated themselves and are joined to fluorinated atoms. Atoms of the third group are unfluorinated. Fluorine atoms have charge within intervals -0.072- -0.080, $e$ ($C_1$) and -0.072- -0.078, $e$ ($D_3$). When fluorination is fully completed in $C_{60}F_{60}$ the charge of both carbon and fluorine atoms is practically constant, equal by absolute value (0.071 and -0.071) and different by sign. As seen from the figure, the real charge distribution has nothing common with the assumptions made in [64]. And who knows, if these charges were taken into account not $D_3$ but $C_1$ molecule would have been the best for the Rietveld refinement.

Much more strong argument in favour of $D_3$ molecule against $C_1$ one is the difference in the total energy, namely -1347.17 and -1293,30 kcal/mol, respectively. The difference is too big to be ignored and seems not leave any doubts in favour of $D_3$. However, accepting $D_3$ for F48 we are facing the next problem. As clearly seen in Fig.2 of Ref. [63], $D_3$ and $S_6$ symmetry dictates a drastic reconstruction of the $C_{60}$ cage that concerns the framing of pentagons which should involves a short double C-C bond. As was shown earlier [66], pentagons in the $C_{60}$ molecule structure are framed by single long C-C bonds only. This finding well correlates with the suggested $6*C_{10}$ structure of $C_{60}$ exhibited by both computational analysis of the molecule chemical reactivity [22-27] and a hypothesis of its formation from mutually bonded carbene chains $C_5$ pointed out about fifteen years ago [67] and actively discussed later [68]. According to this, hexagon-patterned-naphthalene-cores are building blocks of the molecule while pentagons just concern the shape of holes between the blocks. From this viewpoint it is difficult to expect a jump of a short double bond into the pentagon frame since it means the destruction of the naphthalene-core pattern of the molecule. As was shown by the current study, the 6*$C_{10}$ structure did not prevent from the formation of highly symmetric F18 species but did not allow for preserving $C_3$-axis-based symmetry of F36 and F48 species. Geometrical factors given in Table 1 show convincingly that a consequent fluorination of the cage did not require any serious reconstruction of the latter which might result in changing pentagon framing. This means that the main structure factor of the $C_{60}$ cage consisting in the long-single-bond frame of pentagons [68] is conserved during the whole fluorination process.

As for high-symmetry features of $^{19}$F NMR, IR, photoemission, and X-Ray absorption spectra of species F36 and F48, the explanation should be addressed to the actively studied problem of high-symmetry-physical properties of low-symmetry-point-group species [58-60]. A concept of "gray" symmetry which allows for considering the low symmetry as an expansion series involving high-symmetry contribution may give a clear explanation of the observed findings. Thus, coming back to Fig.10 we can see that changing in the charge distribution of F48 caused by the symmetry change is not so drastic. This means that an experimental response based on the charge density will be well similar to both molecular structures and will not be able to distinguish the symmetry difference. Similar should be expected for other physical and chemical properties.

### 5.6. $C_{60}F_{50}$-$C_{60}F_{60}$ adducts

Since $N_D$ has been completely worked out by this stage (Fig.6), one cannot use the pointer and have to proceed with the further fluorination without it. The problem is facilitated by a comparatively small number of empty places that decreases when fluorination proceeds so that the isomer study can be performed just routinely by running over all places one by one. The result of the sorting thus performed is presented in Table 1 and in Fig.11 exhibiting the structure of the species of the lowest total energy. The fluorination process is ended by the formation of $C_{60}F_{60}$ species of $I_h$ symmetry. Since there are no odd electrons in the species, its atoms form an ideal truncated icosahedron structure oppositely to the pristine $C_{60}$ where odd electron conjugation violates the high symmetry lowering it to $C_i$ [60]. When the odd electron behavior is not taken into account, as is the case of the close-shell calculations, the $C_{60}$ symmetry is $I_h$ as well. All above high fluorinated products are thermodynamically stable and could exist.

However, since fluorination is a consequent process their formation becomes energetically nonprofitable at $k \geq 25$ due to positive partial coupling energy (Fig.6) that is why no recording of the species production has been known by now.

**6. Conclusion**

The reactions of fullerene $C_{60}$ with molecular fluorine have been studied using unrestricted broken symmetry HF SCF semiempirical calculations (UBS HF version of the AM1 technique of the CLUSTER-Z1 codes). The calculations are focused on a sequential addition of fluorine atoms to the fullerene cage. A complete family of species $C_{60}F_{2k}$ $k=1,...,30$ has been produced. Based on the effectively-non-paired-electron concept of the selectivity of fullerene molecule chemical activity as well as on a suggested methodology of a computational synthesis of fullerene derivatives, the quantum-chemical computational synthesis of the mentioned derivatives has been performed following the laboratory synthetic pathway of the relevant reactions in the gaseous state. The preferred binding sites for sequential additions are selected by the largest value of atomic chemical susceptibility quantified by the effectively non paired electron fraction $N_{DA}$ on the considered atom $A$. As shown, any addition of fluorine atom causes a remarkable change in the $N_{DA}$ distribution over the $C_{60}$ cage atoms. That is why the synthesis has been performed as a series of predicted sequential steps. Briefly, the synthetic scheme looks like the following. The reaction starts around a pair of fullerene $C_{60}$ atoms with the biggest $N_{DA}$ values. The atoms usually form one of short-length bonds within one of six identical naphthalene-core fragments. When the first adduct $C_{60}F_2$ is formed, the reaction proceeds around a new pair of its fullerene cage atoms with the highest $N_{DA}$ values resulting in the formation of $C_{60}F_4$ adduct. A new $N_{DA}$ map reveals the sites for the next addition step and so on. The reaction stops when all $N_{DA}$ values are fully exhausted. Following these methodology, a complete series of $C_{60}F_{2k}$ species has been synthesized. The obtained results are analyzed from the energetic, symmetry, and the composition abundance viewpoints. A good fitting of the data to experimental findings convincingly proves a creative role of the suggested synthetic methodology in considering fullerene-involved addition reactions of different kinds.

**Acknowledgement**

**Chart 1.** F1 to F8 fluorination. Figures in brackets show the number of the $C_{60}$ cage atom to which the current fluorine atom is attached

| F0-C60 | | F1 (32) | | F2 (31) | | F3 (38) | | F4 (55) | |
|---|---|---|---|---|---|---|---|---|---|
| Atom number | NDA | Atom number | NDA | Atom number | NDA | Atom number | NDA | Atom number | NDA |
| 32 | 0,27072 | 31 | 0.53894 | 38 | 0.29094 | 55 | 0.51918 | 42 | 0.30143 |
| 31 | 0,27077 | 35 | 0.35590 | 20 | 0.29086 | 40 | 0.33620 | 22 | 0.30002 |
| | | 33 | 0.35406 | 55 | 0.29049 | 4 | 0.31575 | 48 | 0.30000 |
| | | 39 | 0.29065 | 18 | 0.29041 | 20 | 0.29535 | 2 | 0.29971 |
| | | 30 | 0.28358 | 10 | 0.26369 | 18 | 0.29026 | 1 | 0.29812 |
| | | 10 | 0.28304 | 40 | 0.26368 | 36 | 0.28537 | 35 | 0.29782 |

| F5 (42) | | F6 (48) | | F7-1(60) | | F8-1(59) | |
|---|---|---|---|---|---|---|---|
| Atom number | NDA | Atom number | NDA | Atom number | NDA | Atom number | NDA |
| 48 | 0.52989 | 60 | 0.34210 | 59 | 0.51853 | 22 | 0.34686 |
| 39 | 0.35424 | 58 | 0.32535 | 57 | 0.37960 | 24 | 0.32989 |
| 22 | 0.34896 | 22 | 0.28518 | 39 | 0.37006 | 54 | 0.29221 |
| *36* | *0.34284* | 1 | 0.28104 | 46 | 0.36312 | 27 | 0.28239 |
| 35 | 0.34226 | 39 | 0.27708 | 60 | 0.34821 | 11 | 0.28026 |
| 49 | 0.34028 | 44 | 0.27568 | 22 | 0.34667 | 36 | 0.27552 |
| | | 35 | 0.27411 | | | 35 | 0.27273 |
| | | 59 | 0.26879 | | | 23 | 0.27116 |
| | | 15 | 0.26840 | | | 28 | 0.27074 |
| | | 2 | 0.26778 | | | 33 | 0.26836 |
| | | 26 | 0.26445 | | | 25 | 0.26625 |
| | | 12 | 0.26395 | | | 6 | 0.26346 |
| | | 57 | 0.26360 | | | 44 | 0.26253 |
| | | 5 | 0.25955 | | | 53 | 0.26195 |
| | | 11 | 0.25790 | | | 29 | 0.25995 |
| | | 27 | 0.25663 | | | 18 | 0.25610 |
| | | 6 | 0.25493 | | | 12 | 0.25594 |

**Chart 2.** Scheme of F8 – F18 fluorination

**F8-1**
**60&57**
**517.162**

**F10-1**
**22&33**
**414.700**

| F12-1 | **F12-2** | F12-3 |
|---|---|---|
| 54&40 | **52&51** | 53&56 |
| 310.952 | **305.920** | 311.329 |

**F14-1**     F14-2
**58&59**     53&56
**198.577**   203.669

F16-1     **F16-2**
53&56     **40&54**
87.656    **86.799**

F18-2    F18-3    **F18-1**
24&23    16&11    **24&23**
-18.287  -8.705   **-24.538**

Table 1. Geometric parameters and total energy of fluorinated fullerene $C_{60}$

|  | F1 | F2 | F3 | F4 | F5 | F6 | F7 | F8 | F10 | F12 | F14 |
|---|---|---|---|---|---|---|---|---|---|---|---|
| $R(C^*-F)^a$, Å | 1,378 | 1,382 | 1,382, 1,378 | 1,382 | 1,382, 1,377 | 1,382 | 1,383-1,381 | 1,384-1,381 | 1,383-1,381 | 1,383-1,381 | 1,387-1,380 |
| $R(C^*-C^*)^b$, Å | $1,52^c$ | 1,61 | $1,52^c$ | 1,62, 1,61 | $1,52^c$ | 1,61 | $1,52^c$ | 1,61-1,60 | 1,61-1,60 | 1,61-1,58 | 1,60-1,58 |
| $\Delta H^d$, kcal/mol | 903,60 | 843,58 | 795,50 | 729,99 | 677,70 | 622,58 | 570,35 | 517,16 | 414,70 | 305,92 | 195,77 |
| Symmetry |  | $C_{2v}$ |  | $C_s$ |  | $C_1$ |  | $C_1$ | $C_1$ | $C_1$ | $C_1$ |

|  | F16 | F18 | F20 | F22 | F24 | F26 | F28 | F30 | F32 | F34 | F36 |
|---|---|---|---|---|---|---|---|---|---|---|---|
| $R(C^*-F)^a$, Å | 1,385-1,378 | 1,384-1,377 | 1,385-1,376 | 1,384-1,375 | 1,385-1,375 | 1,385-1,374 | 1,386-1,376 | 1,385-1,374 | 1,393-1,373 | 1,393-1,376 | 1,402-1,375 |
| $R(C^*-C^*)^b$, Å | 1,60-1,57 | 1,58-1,57 | 1,59-1,57 | 1,59-1,57 | 1,59-1,57 | 1,59-1,57 | 1,58-1,57 | 1,59-1,57 | 1,59-1,57 | 1,59-1,57 | 1,59-1,57 |
| $\Delta H^d$, kcal/mol | 86,80 | -24,54 | -132,87 | -248,07 | -355,61 | -459,33 | -552,60 | -642,47 | -738,80 | -822,14 | -914,95 |
| Symmetry | $C_s$ | $C_{3v}$ | $C_1$ | $C_1$ | $C_1$ | $C_s$ | $C_1$ | $C_1$ | $C_1$ | $C_1$ | $C_1$ |

|  | F38 | F40 | F42 | F44 | F46 | F48 | F50 | F52 | F54 | F56 | F60 |
|---|---|---|---|---|---|---|---|---|---|---|---|
| $R(C^*-F)^a$, Å | 1,402-1,375 | 1,402-1,376 | 1,402-1,381 | 1,402-1,381 | 1,402-1,383 | 1,408-1,387 | 1,408-1,387 | 1,410-1,387 | 1,412-1,387 | 1,412-1,385 | 1,412 |
| $R(C^*-C^*)^b$, Å | 1,59-1,57 | 1,59-1,57 | 1,59-1,57 | 1,59-1,57 | 1,59-1,57 | 1,59-1,57 | 1,59-1,57 | 1,59-1,57 | 1,59-1,57 | 1,59-1,57 | 1,585-1,565 |
| $\Delta H^c$, kcal/mol | -995,77 | -1069,62 | -1120,36 | -1184,68 | -1237,16 | -1293,30 | -1338,08 | -1356,10 | -1371,49 | -1387,42 | -1410,63 |
| Symmetry | $C_1$ | $C_1$ | $C_1$ | $C_1$ | $C_1$ | $C_1$ | $C_1$ | $C_s$ | $C_s$ | $C_s$ | $I_h$ |

[a] C* mark cage atom to which fluorine is added

[b] C*-C* mark a pristine short bond of the cage to which a pair of fluorine atoms is added.

[c] $\Delta H$ is the heat of formation determined as $\Delta H = E_{tot} - \sum_A (E_{elec}^A + EHEAT^A)$. Here $E_{tot} = E_{elec} + E_{nuc}$, while $E_{elec}$ and $E_{nuc}$ are the electron and core energies. $E_{elec}^A$ and $EHEAT^A$ are electron energy and heat of formation of an isolated atom, respectively.

Table 2. $C_{60}F_{48}$ Rietveld refinement [67, 68]

|  | **D3** | **C1** | **S6** |
|---|---|---|---|
| $\chi2$ | 0.1481 | 0.1708 | 0.1870 |
| wRp | 0.0356 | 0.0382 | 0.0399 |
| Rp | 0.0383 | 0.0435 | 0.0438 |
| R (F2) | 0.0758 | 0.0766 | 0.0906 |
| Durbin-Watson d-statistics | 0.6190 | 0.481 | 0.500 |

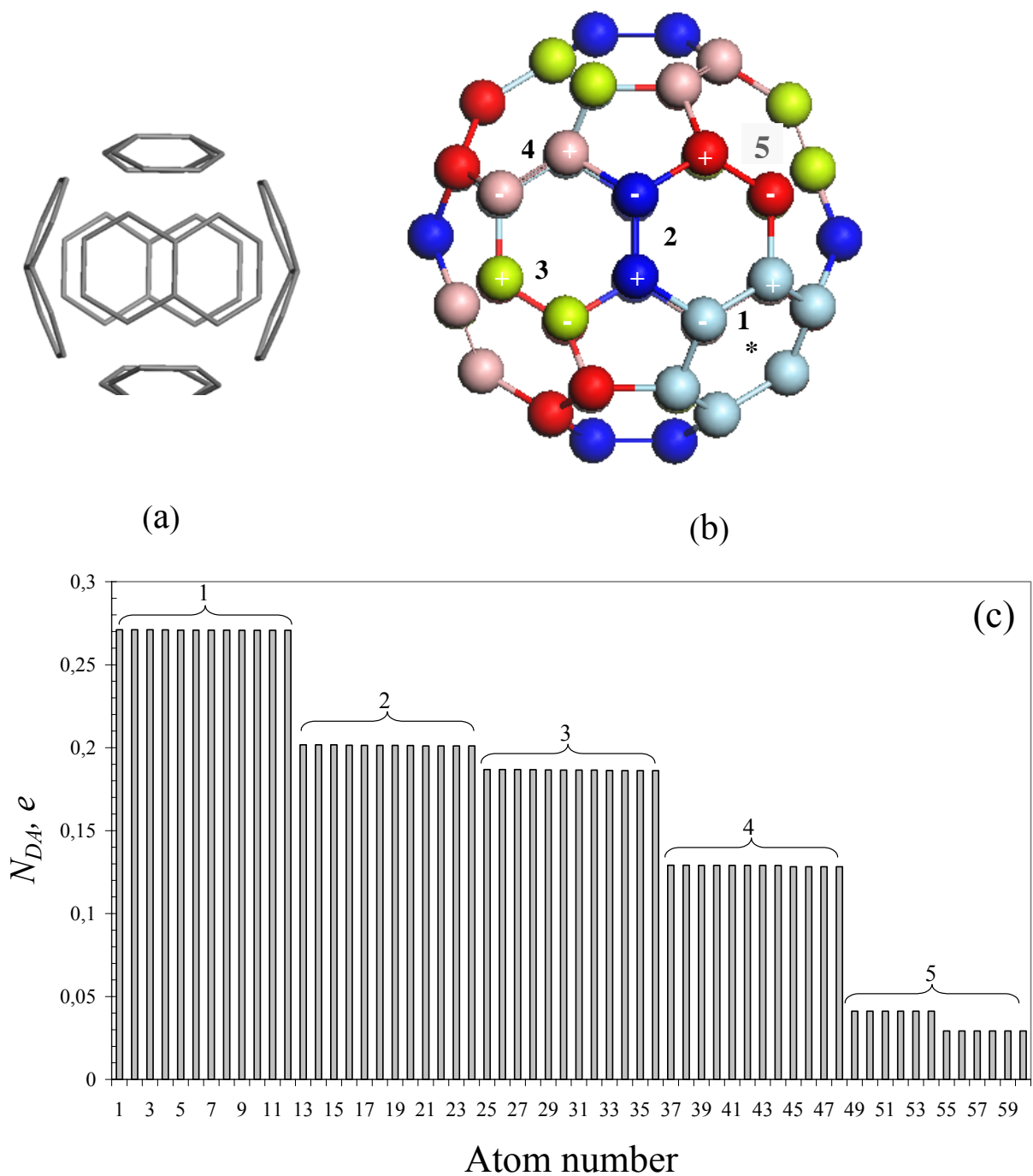

**Fig.1.** Chemical portrait of $C_{60}$ [26]. *a*. 6*$C_{10}$ composition. *b*. Different colors mark atoms with different ACS. Figures point to different atom groups. *c*. ACS map over atoms of $C_{60}$. $N_{DA}$ data are aligned from the biggest to the smallest value.

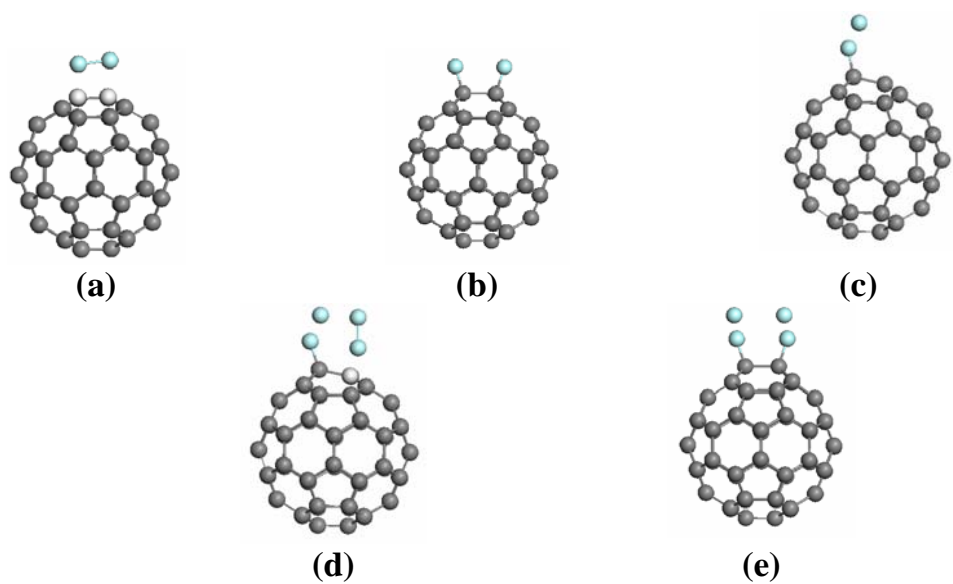

**Fig.2.** Attachment of one (*a-c*) and two (*d, e*) fluorine molecules to the $C_{60}$ cage.  *a*. Starting geometry. Target atoms of the $C_{60}$ cage are shown by light coloring. *b*. Adduct **I** $C_{60}F_2$  *c*. Adduct **II** $C_{60}F_1$ and a free fluorine atom $F_1$; the composition corresponds to the starting geometry in *a*. Starting configuration (*d*) and final adduct (*e*) of the reaction $(C_{60}F_1 + F_1) + F_2$. Target atom of the $C_{60}$ core is shown by light coloring.

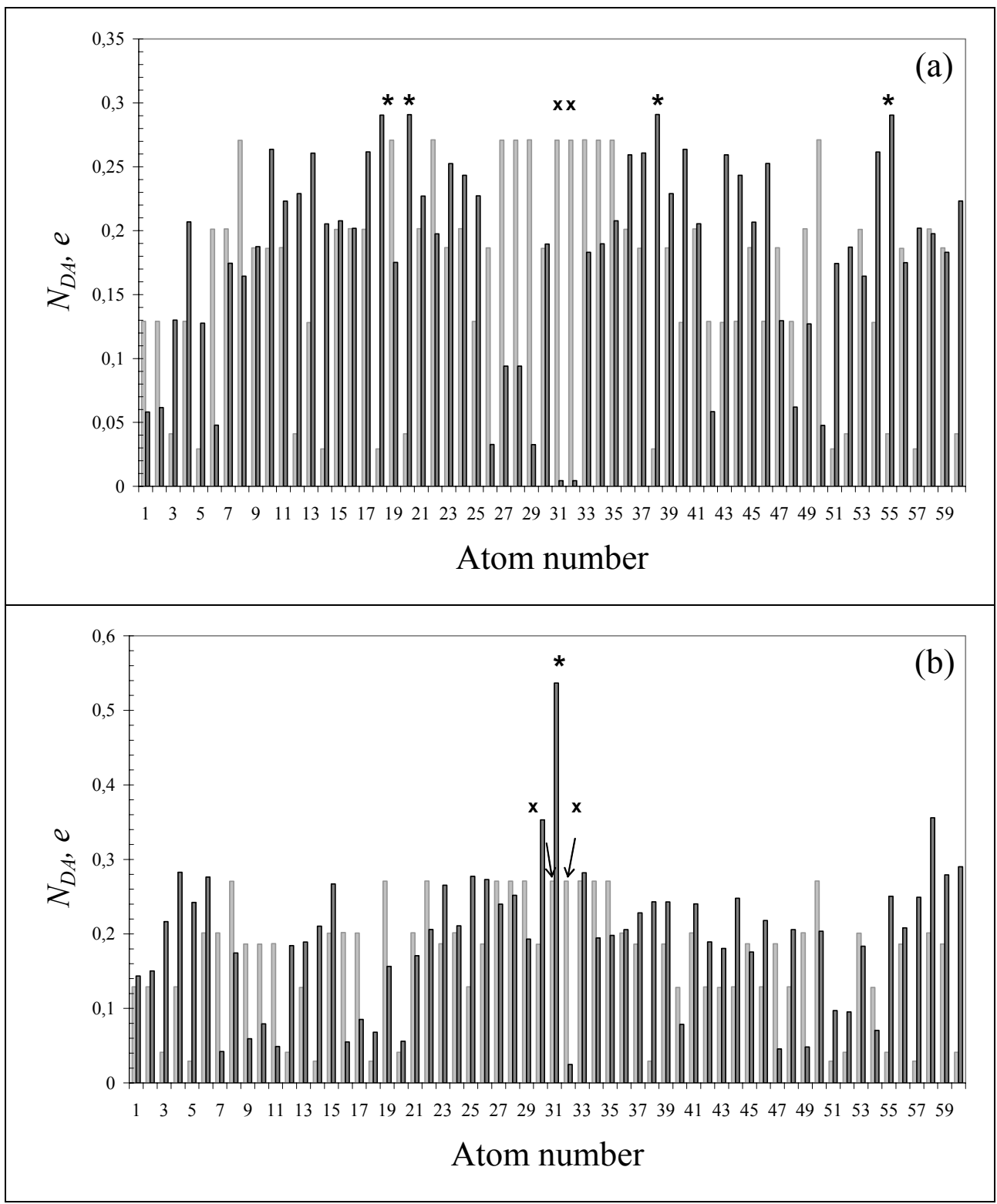

**Fig.3.** ACS map of the $C_{60}$ cage of the adducts $C_{60}F_2$ (*a*) and $C_{60}F_1$(*b*). Light-color bars present the map of the pristine $C_{60}$ molecule. UBS HF singlet state.

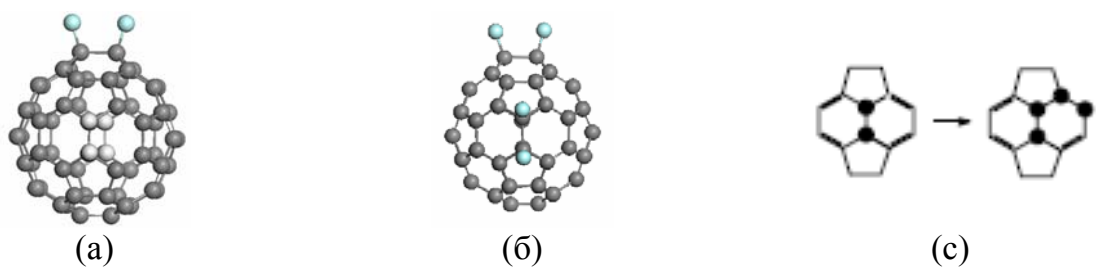

**Fig.4.** Attachment of one fluorine molecule to the $C_{60}F_2$ cage. *a*. Starting geometry. Target atoms of the $C_{60}$ core are shown by light coloring. *b*. Adduct $C_{60}F_4$. *c*. A scheme of contiguous $F_2$ addition [4].

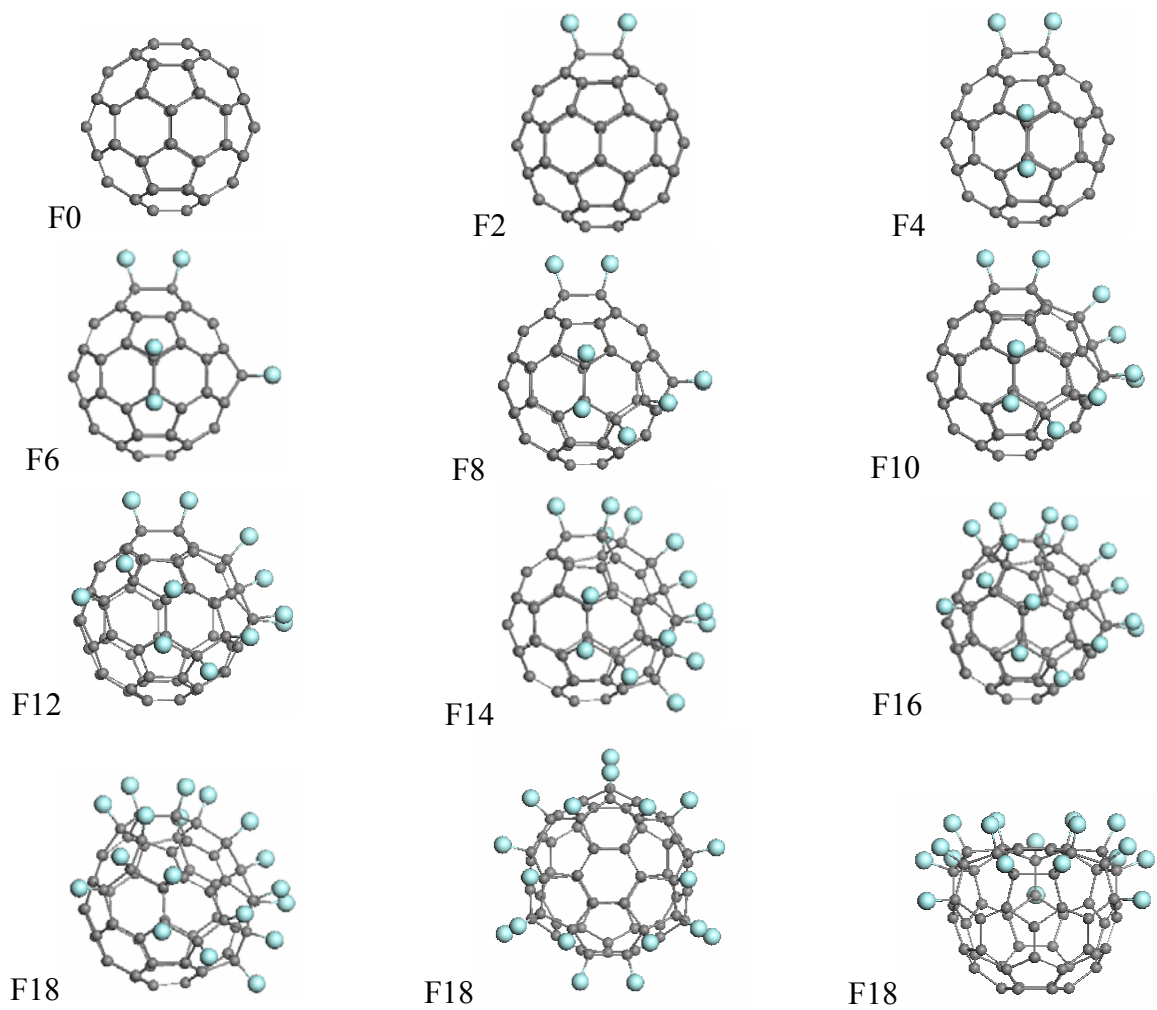

**Fig.5.** Equilibrated structures of $C_{60}F_0$ to $C_{60}F_{18}$ fluorinated fullerenes. UBS HF singlet state.

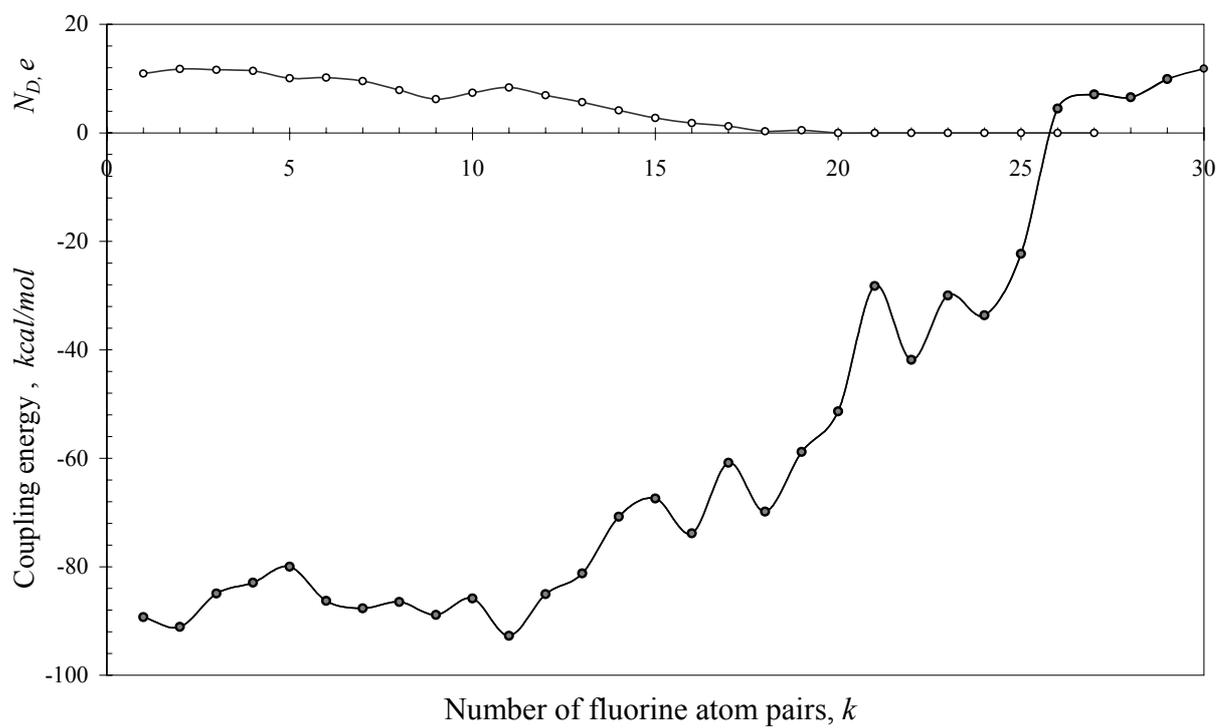

**Fig.6.** Evolution of the $N_D$ value and coupling energy $E_{cpl}$ with the number of fluorine atom pairs $k$.

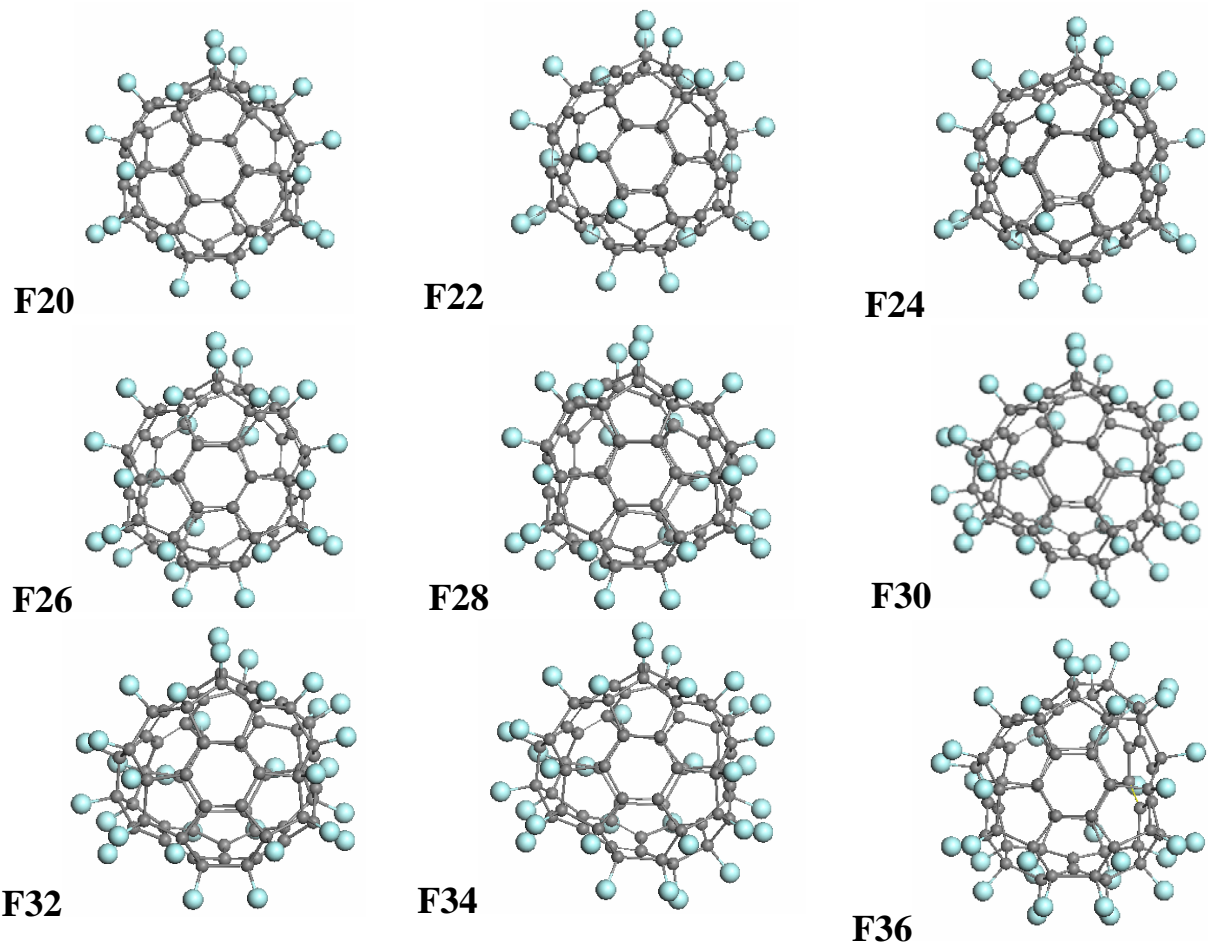

**Fig.7.** Equilibrated structures of $C_{60}F_{20}$ to $C_{60}F_{36}$ fluorinated fullerenes. UBS HF singlet state.

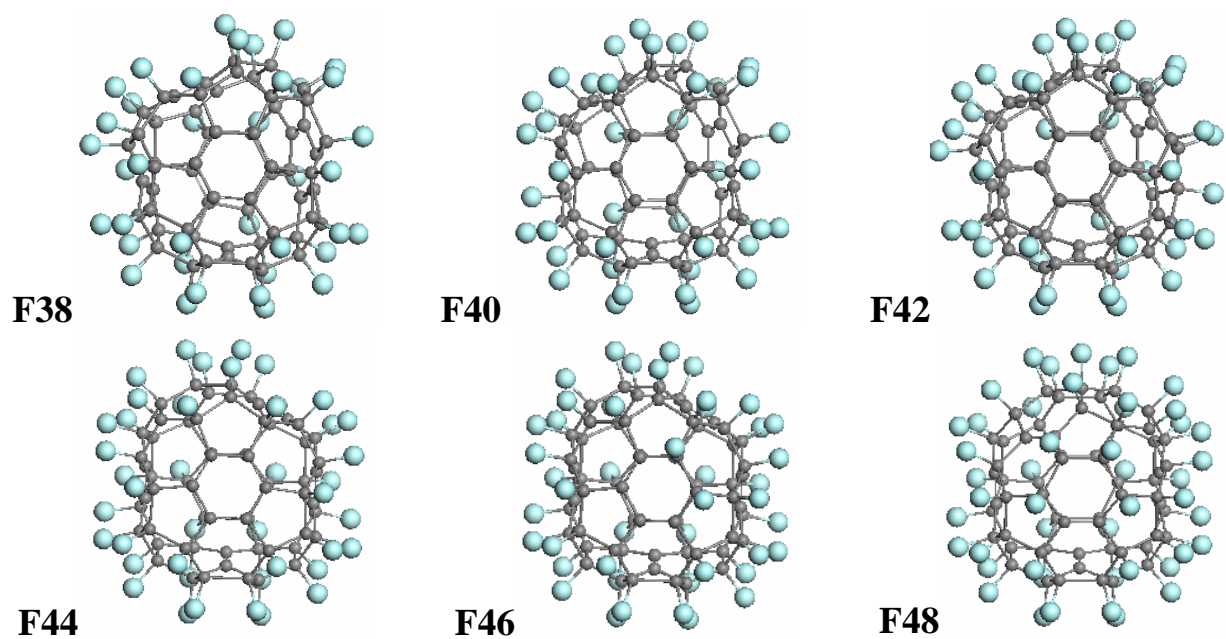

**Fig.8.** Equilibrated structures of $C_{60}F_{38}$ to $C_{60}F_{48}$ fluorinated fullerenes. UBS HF singlet state.

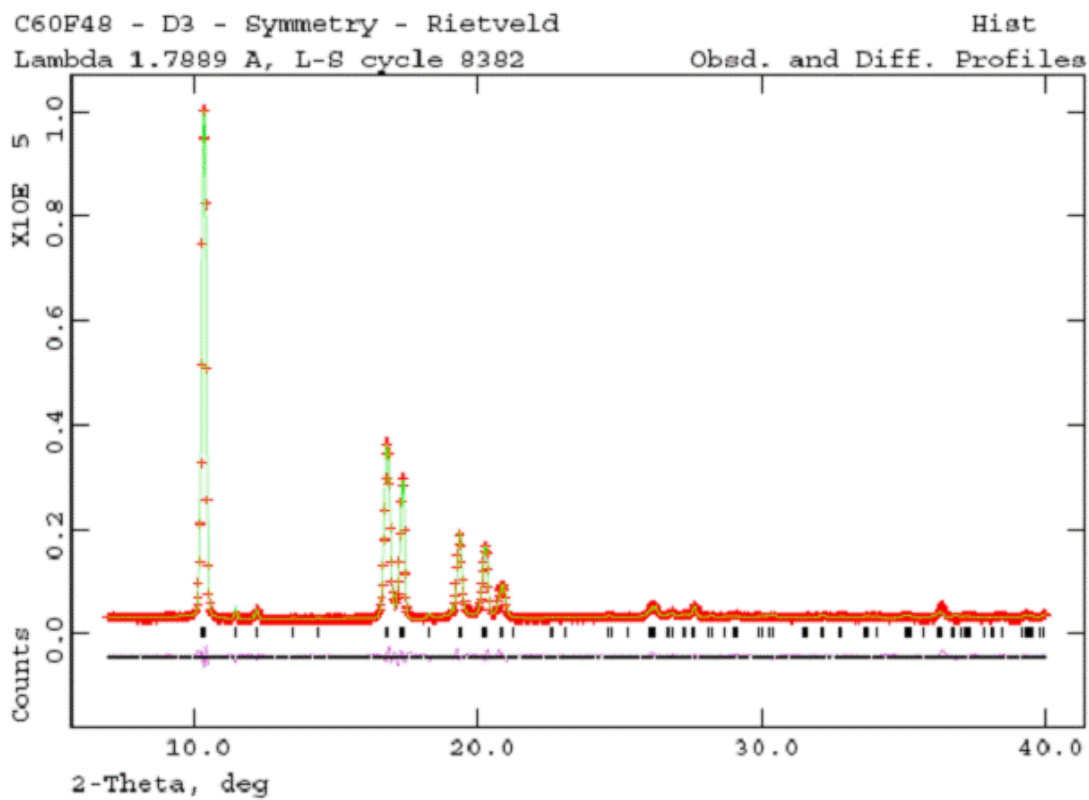

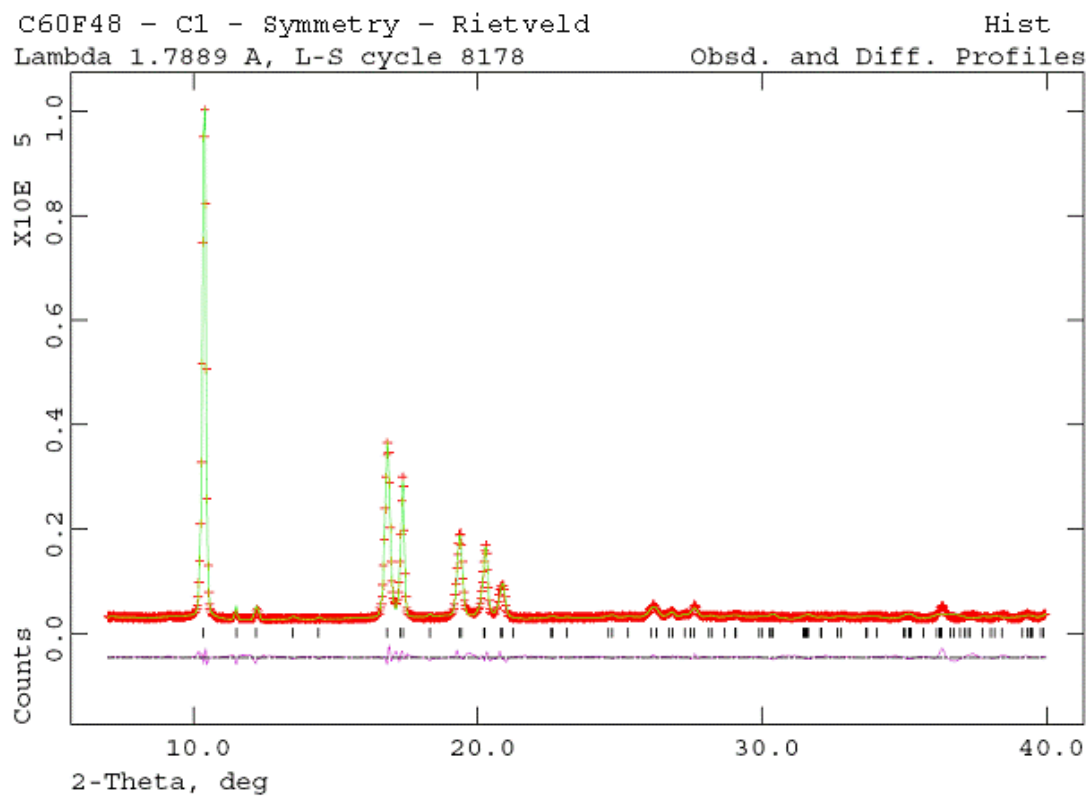

**Fig.9.** Plots of the best GSAS-based Rietveld refinements associated with $D_3$ [64] (top) and C1 [65] (bottom) molecular models courtesy by Dr. R.Papoular, Leon Brillouin Laboratory, CEN-Saclay

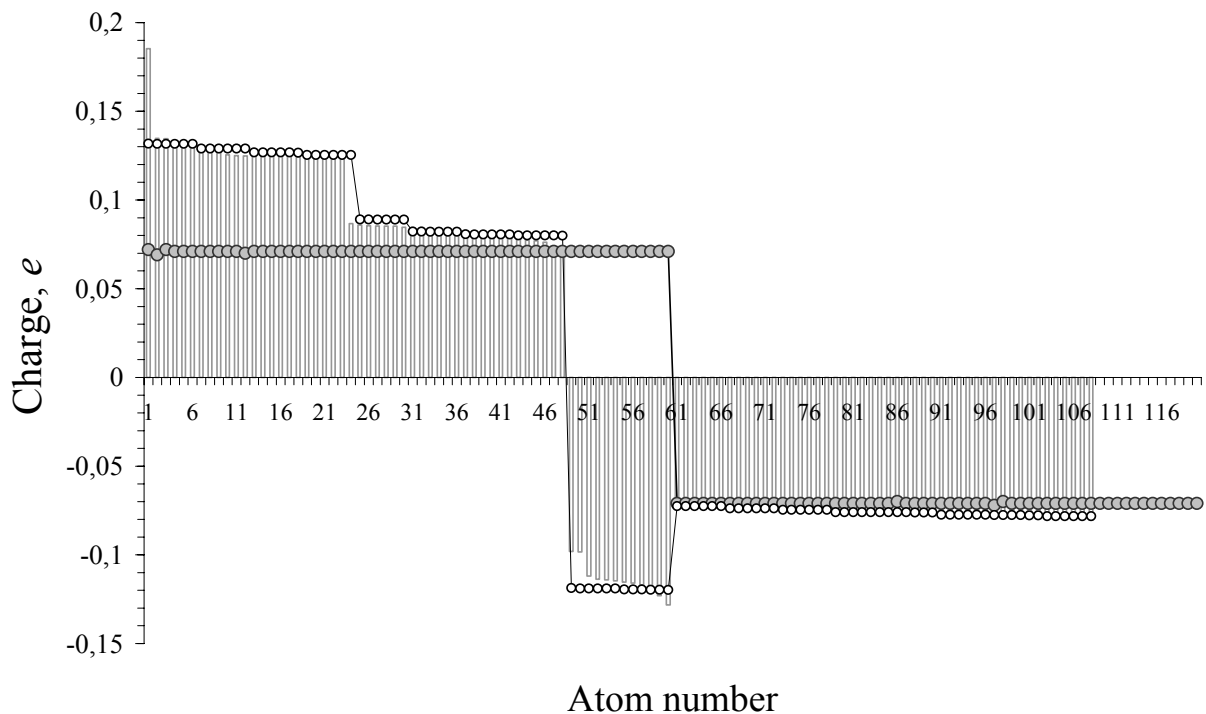

**Fig.10.** Charge distribution over atoms of $C_{60}F_{48}$ (histogram- $C_1$ molecule and curve with empty dots – $D_3$ molecule) and $C_{60}F_{60}$ (curve with filled dotes) species. UBS HF singlet state

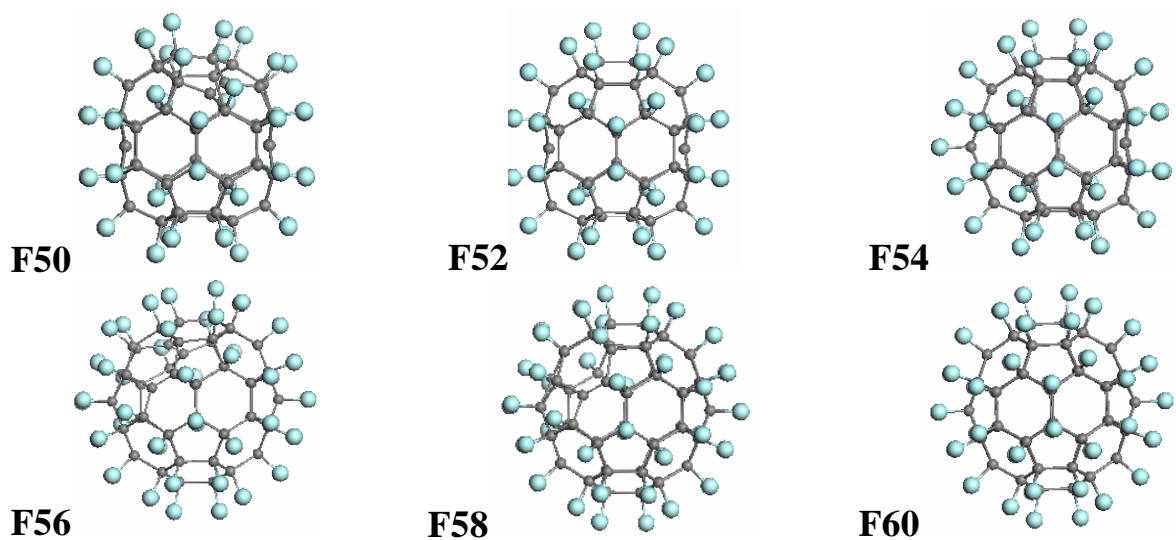

**Fig.11.** Equilibrated structures of $C_{60}F_{50}$ to $C_{60}F_{60}$ fluorinated fullerenes. UBS HF singlet state.